\renewcommand{\@maketitle}{\newpage
 \null
 \vskip 2em                 
 \begin{center}
  {\Large \bf \@title \par}     
  \vskip 1.5em                
  {\large                        
   \lineskip .5em           
   \begin{tabular}[t]{c}\@author
   \end{tabular}\par}
  \vskip 1em              
  {\large \@date}           
\end{center}
 \par
 \vskip 2.5em}                

\renewcommand{\abstract}{
\if@twocolumn
\section*{Abstract}
\else \normalsize
{\bf Abstract}
\fi
\par}

\def\endabstract{\if@twocolumn\else
\fi
\vfill\eject}

\def\section{\setcounter{equation}{0}
\@startsection {section}{1}{\z@}{-3.5ex plus -1ex minus
 -.2ex}{2.3ex plus .2ex}{\large\bf}}
\documentstyle [12pt] {article}
\documentstyle{peer}

\newcommand{\bye}{\end{document}}

\newcommand{\baufg}{\begin{description}}
\newcommand{\eaufg}{\end{description}}
\newcommand{\bteilaufg}{\begin{description}}
\newcommand{\eteilaufg}{\end{description}}

\newcommand{\be}{\begin{equation}}
\newcommand{\ee}{\end{equation}}
\newcommand{\bes}{\begin{eqnarray}}
\newcommand{\ees}{\end{eqnarray}}
\newcommand{\bma}{\left( \begin {array}}
\newcommand{\ema}{\end {array} \right)}
\newcommand{\pslash}{\kern 0.2 em p\kern -0.45em /}
\newcommand{\dslash}{\kern 0.2 em \delta\kern -0.45em /}
\newcommand{\sla}[1]{\kern 0.2 em #1\kern -0.45em /}

\newcommand{\bt}{\begin{tabbing}
            \hskip 7.1 true cm \=\hskip 7.1 true cm \kill}
\newcommand{\et}{\end{tabbing}}
\newcommand{\bfig}{\begin{figure}}
\newcommand{\efig}{\end{figure}}

\begin{document}
\input prepr.tex
\def \fhat{\hat{F}}
\def \tr{\rm Tr}
\def \Re{\rm Re}
\def \MeV{\rm MeV}
\def \element{\rm element} 
\def \twiggle{\tilde}
\newcommand{\vx}{\mbox{$\vec{x}$}}
\newcommand{\vy}{\mbox{$\vec{y}$}}
\def \v0{\vec{0}}
\newcommand{\lae}{\raisebox{-0.3ex}
{$\renewcommand{\arraystretch}{0.4}\begin{array}{c} < \\ \sim \end{array}$}}
\newcommand{\gae}{\raisebox{-0.3ex}
{$\renewcommand{\arraystretch}{0.4}\begin{array}{c} > \\ \sim \end{array}$}}
\begin{flushright}
CERN-TH 6692/92\\
PSI-PR-92-27\\
WUB 92-34\\
\end{flushright}

\begin{center}
{\bf
\Large
The Static Approximation of Heavy-Light Quark-Systems \\
- A Systematic Lattice Study\footnote{work
supported in part by DFG grant Schi 257/3-1.} - \\}
\end{center}
\normalsize
\medskip

\begin{center}
C. Alexandrou$^{a}$, S. G\"usken$^{b}$,  F. Jegerlehner$^a$, K. Schilling$^b$,
R. Sommer$^c$\footnote{
     {\it on leave of absence from University of Wuppertal.}}\\
\end{center}

$^a$ {\it PSI, CH-5232 Villigen, Switzerland\\}
$^b$ {\it Physics Department, University of Wuppertal
              D-5600 Wuppertal 1, Fed. Rep. \\  Germany\\}
$^c$ {\it CERN, Theory Division, CH-1211 Geneva 23, Switzerland}\\

\centerline{{\bf Abstract}}
\noindent

We present a study of {\it finite} $a$ and {\it volume} effects
of the leptonic decay constant $f$ of heavy pseudoscalar mesons in
the static approximation.
This study is performed on a number of lattices
at $\beta=$ 5.74,~6.0 and 6.26 covering sizes from about 0.7~$fm$ to 2~$fm$.
We confirm that beyond 1.5~$fm$ the volume dependence is negligible.
By carefully analysing results obtained using different trial wave functions
for the heavy meson we find no dependence on the smearing.
We also give results for the mass difference of the scalar - pseudoscalar and
the $\Lambda _{b}$ - pseudoscalar. Using the mass of the pseudoscalar meson we
estimate the distance of string breaking in the static quark potential.
\noindent

\newpage
\section{Introduction}

There is continued interest in applying the
infinite mass  effective theory (IMET) \cite{harvard} to compute
properties of heavy-light quark systems in the range
of the b- and c- quarks. However the issue of the
domain of its validity is still unsettled: it most likely
depends on the particular observables under consideration.
Lattice methods offer a laboratory to establish beyond which mass
 the
infinite mass limit could be regarded as a good approximation.
The lattice formulation of IMET, originally suggested
by Eichten \cite{Eichten}, has been
studied subsequently by a number of authors [3-9].
There are strong indications, that the infinite mass limit
of the  leptonic decay constant $f$ of the pseudoscalar heavy-light
meson needs large corrections to be applicable in the D-meson region.
In order to reach a definite conclusion on $f$ and
 other quantities such as the scalar-pseudoscalar mass splitting
$\Delta_S$
and the $\Lambda_b$ - pseudoscalar mass difference $\Delta_{\Lambda}$, however,
one has to gain better control on
various systematic effects of the
infinite mass lattice results.
There are a number of issues to be addressed
such as constraints of finite
lattice resolution and extension, finite $u$ and $d$ quark mass and
the necessity of renormalisation.
In addition one has to bear in mind,  that before one can  extract any
reliable information, one has to achieve ground state
enhancement by suitable smearing techniques \cite{fb1,fb2,Step}.

In this paper, we  intend to {\rm consolidate} previous
static lattice results on $f$ \cite{fb1,fb2}
with respect
to the error sources  mentioned above and to explore the feasibility of
studying $\Delta_S$ and $\Delta_{\Lambda} $. For that purpose,
we present in the following
 a fairly comprehensive analysis, performed
on lattices ranging in size from about 0.7 to 2 $fm$
at $\beta=$~5.74,~6.0 and 6.26. Particular emphasis will be  given to
the pitfalls of ground state projection.
We  will show that we can safely identify the ground state
contributions in the b-quark mesons when smeared interpolating fields
are used for the mesons.
Using several different interpolating fields we find no dependence
of the final result on the type of wave function used. This is what
is to be expected and it is contrary to results found in ref.~\cite{Jap}.

In order to connect
the lattice results to physics, it is important to
address
finite $a$ and finite size effects as well as
renormalisation. We find that the volume effects
are negligible beyond lattice sizes of 1.5~$fm$. Finite $a$ effects
 turn out to be small when choosing
the string tension as the appropriate
mass scale. In this way, the most serious source of uncertainty on
$f$ remains the perturbative determination of
its  renormalisation factor $Z$. Of course all statements that we make here
are true within the quenched approximation. We have no way of giving a
reliable estimate of the error due to quenching.

The paper is organized as follows: Section 2 will deal with
the smearing technique for  ground state enhancement using gauge invariant
 wave
functions.
Section 3 contains the lattice results on the pseudoscalar
mass and leptonic decay constant as well as the mass splittings of
scalar-pseudoscalar,
$\Delta_S$, and $\Lambda_b$-pseudoscalar, $\Delta_{\Lambda}$.
Their connection  to physical quantities is discussed in section 4.
The reader not interested in lattice technicalities may go directly to
section~4.
In section~5 we discuss  the availability of the
physics of string breaking
in the environment of quenched computations and give an estimate of
the string breaking distance $R_b$ obtained in the quenched approximation.
The parameters  and technical details
characteristic to our simulations are collected for convenience
in appendix A.

\section{Smearing Techniques}
In this work we consider correlation functions
of interpolating fields for hadrons that consist of light quarks ($l$) and one
heavy quark ($h$). The heavy quark is treated in the static
 approximation with its propagator given by\cite{Eichten,Eich81}
\bes
S_h(x;y)= \delta_{\vx,\vy} ~~ \left \{  \right. & \left.
\Theta (x_4-y_4) \; W^{\dagger}(\vx;y_4,x_4)  \gamma^+
\right.                                           \label{prop}
\\ \nonumber  \left. +  \right. &  \left.
\Theta (y_4-x_4) \; W(\vx;x_4,y_4)  \gamma^-
 \right \}, \\
\gamma^{\pm }=(1 \pm \gamma_4 )/2 \quad . \nonumber
\ees
Here $ W(\vx;x_4,y_4)=\prod_{t=x_4}^{y_4-a}~U_4(\vx,t)$ is the
gauge parallel
transporter from point $(\vx,x_4)$ to $(\vx,y_4-a)$.
In eq.(\ref{prop}), the exponential prefactor $\exp(-|x_4-y_4|~m_h)$ has been
dropped, since this corresponds only to a common shift of all energy levels
by the bare mass of the heavy quark.

In order to connect to physical i.e. renormalisation group invariant quantities
within the static approximation, we have to go beyond the formal expression
of eq.(\ref{prop}).
This amounts to taking into account the renormalisation of currents
and the effect of the self energy of a static quark which
in the continuum limit is divergent.
We will discuss this issue in more detail in section 4.1.
Given the renormalisation constant of the axial current, $Z_{stat}$,
the physical decay constant $f$
is calculated through the
lattice matrix element
\be
<0|{\cal{M}}^{loc,loc}_{\gamma_4\gamma_5}(0)|P> = Z_{stat}^{-1}~f
 \sqrt{M_{P}/2}~~ a^{3/2}
\quad .
\label{deccon}
\ee
In this equation, all quantities are understood to be in the
static approximation.
For later use  we define
\be
\fhat = f \sqrt{M_P}
 \left (\frac{\alpha_s(e^{-2/3}M_P)}{\alpha_s(e^{-2/3}M_B)} \right
)^{6/25}
\quad .
\label{fhat}
\ee
The matrix element $<0|{\cal{M}}^{loc,loc}_{\gamma_4\gamma_5}(0)|P>$ is
independent of the mass of the heavy quark, $m_h$.
The logarithmic
mass dependence of $f \sqrt{M_P}$ enters through  the mass dependence
of $Z_{stat}$. This logarithmic dependence is cancelled in the definition of
$\fhat$, which is the proper quantity that has a finite limit as
$m_h \rightarrow \infty$.
The field  ${\cal M}^{loc,loc}_{\gamma_4\gamma_5}(\vx,t)$ is the time component
of the local axial vector current in the general notation
\be
{\cal M}^{I,J}_{\Gamma}(\vx,t) = \bar{h}^I (\vx,t)
{}~\Gamma ~l^J(\vx,t)\quad ,      \label{meson}
\ee
where the indices $I, J$ denote smearing local fields with
trial wave functions of type
\be
l^J(\vx,t) = \sum_{\vy} \Phi ^J(\vx,\vy;{\cal U}(t))~l(\vy,t) \quad ,
\label{smear}
\ee
and $\Gamma=\gamma_4\gamma_5$ in eq.~(\ref{deccon}).
$\Phi ^J(\vx,\vy;{\cal U}(t))$ depends in a gauge invariant way on the
link variables  denoted by ${\cal U}(t)$.
The case of the local interpolating field that
appears in the definition of
the decay constant in eq.(\ref{deccon}) is given by
$$
\Phi ^{loc} (\vx,\vy;{\cal U}(t) ) = \delta_{\vx\vy}~.
$$
In order to calculate the matrix element required in eq.(\ref{deccon}) one
 starts
from the  meson--meson correlation function
\be
C^{I,J;K,L}_\Gamma(t) = \sum_{\vx} <  {\cal M}^{I,J}_{\Gamma}(\vx,t)~ [ {\cal
M}
^{K,L}_{\Gamma}(\vec{0},0)]^\dagger >  \quad .    \label{correla}
\ee
Due to the positivity of the transfer matrix~\cite{Transf}, this has the
general
representation
\be
C^{I,J;K,L}_{\Gamma}(t) = \sum_{n \geq 1}
 <0|  {\cal M}^{I,J}_{\Gamma}(\vec{0},0) | n, \gamma>< n, \gamma| {\cal
 M}^{K,L}_{\Gamma}(\vec{0},0)
]^\dagger |0>
             \exp( - \tilde{M}_n(\gamma) t ) \label{spect} \quad .
\ee
Note that we distinguish between     the ``masses'' $\tilde{M}_n$
in the static approximation, that appear in the above formula and the
physical mass of the meson (e.g. $M_P$ ) as used in eq.(\ref{deccon}).
Eq. (\ref{spect}) gives the formula appropriate for an
infinitely large temporal lattice, since
for the time extent of about $4$ $fm$ used in our simulations,
corrections to the infinite limit are expected to be negligible.
We have also assumed in this decomposition, that the trial wave functions
$\Phi ^J(\vx,\vy;{\cal U}(t))$
do not introduce angular momentum and hence the intermediate
meson states $| n, \gamma>$ are uniquely labelled  by the spin/parity quantum
 numbers characteristic to  $\Gamma$ and the excitation level $n$.
Due to the  additional spin symmetry of IMET~\cite{spin}
we only need to distinguish between pseudoscalar
P ($\gamma=\gamma_5$) and scalar S ($\gamma=1$) particles, the vector being
degenerate to the pseudoscalar.

For sufficiently large time separation $t$, all contributions
are exponentially suppressed with respect to  the ground state ($n=1$) in
eq.(\ref{spect}).
Therefore, the matrix element defining $f$ in
eq.(\ref{deccon}) can be obtained by
choosing $I,\dots,L=loc$ referred to as ``local-
local" correlation function.
In the following, we will use the shorthand notations
$|1, \gamma_5>=|P>$, $M_1(\gamma_5)= M_P$ and  $M_1(1)= M_S$.

Ground state dominance of
the correlation function is signalled
by the occurrence of a plateau, {\rm i.e.}
the `local mass'
\be
\mu^{I,J;K,L}_{\Gamma}(t) =
 {\log}(C^{I,J;K,L}_{\Gamma}(t)/C^{I,J;K,L}_{\Gamma}(t-a)) \quad .
 \label{mass}
\ee
is t-independent for a range of $t$.
So far, it has been found impossible to attain  such a plateau
within the static approximation,
before the signal
is lost in the statistical noise.
The situation of our purely local data from lattice D2a
(the notation for  our lattices being defined
in table~1) is
depicted by the diamonds in fig.~1.
Given the statistical errors
it is again not possible  to establish
the existence of a plateau, up to time separation of
$t=10a$ (or $t\simeq 3.6 GeV^{-1}$).

The situation is greatly improved by using
appropriate  smearing techniques~\cite{Step}.
Translation invariance and the hermiticity of $\Phi$ imply the
symmetry relations
for the expectation values
\be
C^{I,loc;J,loc}_\Gamma (t) = C^{loc,I;loc,J}_\Gamma (t) \quad .
\ee
In other words we can choose whether to
smear the light  or the static quark in the correlator,
without change in the expectation values.
Moreover, we have
\be
C^{I,loc;J,loc}_\Gamma (t) = C^{J,loc;I,loc}_\Gamma (t)  \quad .
\ee

The statistical errors of these quantities, however,  do not at all
obey such identities!
We demonstrate this fact for the second of these equations in
fig.~2 where we show a comparison of the `local masses'
obtained from $C_{\gamma_4\gamma_5}^{loc,loc;I,loc}(t)$ and
$C_{\gamma_4\gamma_5}^{loc,loc;I,loc}(t)
+ C_{\gamma_4\gamma_5}^{loc,I;loc,loc}(t)$ for gaussian smearing.
There is a dramatic difference in errors between the
two cases. So it is much more advantageous to smear the
static quark source than to apply smearing to the
static sink\footnote{This observation has been made  already in
ref. \cite{Eichca}.}.
The appendix B  presents arguments for this empirical
error behavior.
The difference in error bars shown in fig.~2 amounts to
a significant improvement in our ability  to localize the plateau\footnote
{It turns
out that for the final determination of the decay constant, however,
the statistical error is very similar in the two cases.}.
Smearing both heavy and light quarks on
lattice C4 at $\beta=6.0$, we could not do better.

In order to construct the purely local quantity
$<0|{\cal{M}}^{loc,loc}_{\gamma_4\gamma_5}(0)|P>$
from the smeared correlation function
$C^{I,J;I,J}_{\gamma_4\gamma_5}(t)$ and
$C^{loc,loc;I,J}_{\gamma_4\gamma_5}(t)$, one
has first to establish their plateaus as described in eq.(\ref{mass}).
We fit data showing ground state dominance to the one-exponential
expressions
\bes
C^{I,J;I,J}_{\gamma_4\gamma_5}(t) & \simeq & |<0|
{\cal M}^{I,J}_{\gamma_4\gamma_5}(\vec{0},0) | P> |^2
             \exp( - M_{P} t ) \nonumber \\
&  =& A \exp(- M_{P} t)\label{fitss}\\
C^{loc,loc;I,J}_{\gamma_4\gamma_5}(t) &\simeq & <0|
{\cal M}^{loc,loc}_{\gamma_4\gamma_5}(\vec{0},0) | P>
<P| {\cal M}^{I,J}_{\gamma_4\gamma_5}(\vec{0},0)
]^\dagger |0>
             \exp( - M_{P} t ) \nonumber \\
& =& B \exp(- M_{P} t)\quad  \label{fitls}
\ees
and  retrieve the local matrix element in form of the ratio
$B/\sqrt{A}$.

Clearly the trial wave functions $\Phi^I$ and $\Phi^J$ have to be constructed
 such that
the excited state contributions are sufficiently suppressed so that
the one-exponential behavior in the previous equations can be observed
before the statistical noise dominates the signal for both correlators
given in (\ref{fitss}) and (\ref{fitls}).

To achieve this goal, we have considered three types of wave
 functions\cite{Step}:
\begin{enumerate}

\item Gaussian:\\
$$\Phi ^G(\vx,\vy;{\cal U}(t)) = ( 1 + \alpha
     H )^n(\vx,\vy;{\cal U}(t))$$

with the hopping matrix
$$     H(\vx,\vy;{\cal U}(t)) = \sum^3_{i=1}
            (  U_i(\vx,t)  \delta_{\vx,\vy-\hat{i}}
  +  U_i^{\dagger}(\vx - \hat{i},t)  \delta_{\vx,\vy+\hat{i}} )$$
that contains the optimisation parameters $\alpha$ and $n$.
\item Exponential:\\
$$\Phi ^E(\vx,\vy;{\cal U}(t)) = ( K \delta_{\vx\vy} -
     H(\vx,\vy;{\cal U}(t)) )^{-1}$$
with the optimisation parameter $K$.
\item Combinations: \\
$$\omega ~\Phi ^{loc}(\vx,\vy;{\cal U}(t)) + \Phi ^G(\vx,\vy;{\cal U}(t)).
$$
\end{enumerate}

Let us briefly discuss some features of these wave functions:\footnote{
Additional aspects are discussed in an early investigation~\cite{Step}.}

All wave functions are defined local in time which means that they involve
gauge fields only in one time slice and are gauge covariant.
The first property is needed in order that eq. \ref{spect} holds.
The second one insures that no problems due to gauge fixing occur.
Such problems could arise, for example, due to noise originating from
incomplete gauge fixing. In addition, the necessity to work in the Coulomb
gauge for all other wave functions that have been suggested \cite{smear}
means that a significant amount of numerical effort is spent in the
gauge fixing procedure.
The gaussian wave function given above is therefore
comparatively cheap to compute.

The exponential wave function needs the calculation of a 3-dimensional scalar
propagator through the solution of a linear equation. We refer to it as
an exponential wave function because the $\vx - \vy$ dependence
in the non-interacting case ($U_i(\vx,t)=1$) is of Yukawa-type.
This particular wave function was found~\cite{fb1} to be very effective
in projecting onto the ground state at relatively small $\beta$ ($\beta\simeq
 5.7)$.
In our new simulations, we have switched to using the gaussian
wave function (this notation originates again from its
dependence on $\vx - \vy$ in the free case), because it is numerically
much cheaper. As we will demonstrate
in the following, the two wave functions are equally
good in projecting onto the ground state around $\beta=6.0$
and for $\beta=6.26$ the gaussian wave
function achieves very early ground state dominance. Although $\Phi^E$ has not
been tried out at this $\beta$ value, we expect $\Phi^G$ to be superior in
ground state projection.

In our previous calculation we tuned the parameter(s) in the wave function
to obtain early plateaus in the smeared-smeared {\it and}
 local-smeared  local masses.
Defining
\be
r_2^2 = < \frac {\sum_{\vx} \vx^2 {\rm Tr} [\Phi ^J(\vx,0) (\Phi
 ^{J}(\vx,0))^{\dagger}
]}
               {\sum_{\vx}      {\rm Tr} [\Phi ^J(\vx,0) (\Phi
 ^{J}(\vx,0))^{\dagger}
]} >  \quad  \label{rms}
\ee
the optimisation of the wave functions coincided
with an r.m.s. radius
$r_2$ of approximately 0.3 fm. This is quite a reasonable size for a hadronic
 wave function.

\section{Signals from the Lattice}
\subsection{Pseudoscalar Mass and Decay Constant}
As we already pointed out, smearing of the quark fields is crucial for
filtering the ground state before noise sets in.
Needless to say, the results must be independent of the
details of the applied smearing.
This issue has been addressed already in the first calculations of $\fhat$
in the static approximation~\cite{fb1,Allton}, where different
wave functions were reported to render  consistent results.
More recently ref.~\cite{Jap} supposedly
revealed a systematic dependence of $\fhat$ on
the wave function size, particularly at
larger values of
$\beta$.
In order to settle this question,
 we
performed a
 careful analysis on wave function dependence at $\beta=6.26$.
We used a $18^3 \times 48$ lattice (denoted by $D2a$
in table~1), and six different wave functions, whose
parameters are listed in table 2.
The first three wave functions
are constructed using a combination of a local and a gaussian wave function
with $n=100,\alpha=4$. The remaining
three represent pure gaussian smearing.
To convey  a physical
idea of the involved sizes we quote  the r.m.s radii $r_2$ (defined
in eq.(\ref{rms})) as well as the radii $r_1$ (defined as in eq.(\ref{rms}) but
with   $\vec{x}^2$ replaced by $|\vec{x}|$).
As can be seen from table 2, $r_2$ is varied within $3.9a \leq r_2 \leq 6.5a$.

The impact of smearing on
the t-dependent masses, as derived from
$C_{\gamma_4\gamma_5}^{I,loc;J,loc}$  (``smeared - smeared'')
is demonstrated in fig.1.
We can clearly ascertain  a universal  plateau, independent of the
particular  wave function chosen.
Approximate ground state
dominance sets in  as early as  $t = 2a$.
Fits to eq.(\ref{fitss}) in the t-range of the plateau
yield consistent mass values, as can be seen from table~2\footnote{
Fig.~1 shows quite a strong scatter in the data at intermediate and
large values of $t$.
We investigated this in some detail:
The error estimates of the t-dependent masses were calculated using
the covariance matrix.
One may suspect that
the off diagonal
elements of the covariance matrix in themselves are not estimated accurately
enough,
leading to unreliable error estimates in the t-dependent masses.
To check this suspicion, we estimated the {\it error of } the error with the
jackknife procedure and found it to be only of the order of 15\%, in agreement
with a gaussian distribution. Furthermore no signs of
significant
 autocorrelations were revealed by the data (using binning). We conclude
that the reason for the scatter in fig. 1 is that
correlations over more than one time-slice are not strong.
This fact actually improves the errors for the mass
from the smeared-smeared correlators: different patches of $t$
contribute independent information.
}.

Since the smeared-smeared correlation function  is convex
(cf. eq.~(\ref{spect})), we were able to
identify its  ground state plateau rather
unambiguously.
The evaluation of local-smeared correlation functions, on the other hand,
poses a more serious problem, as they
do not
share this convexity property.
We show their t-dependent masses in fig. 3: without prior knowledge of
the height of the plateau (and in particular with the larger error bars
in the t-dependent masses obtained in ref.~\cite{Jap}),
one is liable to  misidentify the position
of the plateau and thus to end up with  wave function dependent results.
It is therefore important to ensure  that the plateau
has the same height  for both local-smeared and smeared-smeared correlators.
For this purpose, we have inserted into fig.~1 and 3 the plateau as obtained
from a fit to smeared-smeared correlators.
(using the best  wave function with $\alpha=4,~n=100$).
Although the t-dependent masses depend
substantially on the underlying wave functions, they
finally end up  in the same plateau!

So it is obvious that the evaluation of smeared correlators
has its pitfalls.
An appropriate  procedure to determine
$\fhat$ is to first determine the  ground state mass from
the smeared-smeared correlator and then to  inject it into a  constrained fit
of the local-smeared correlator.
These fits, done in the region of the respective plateaus
yield consistent results. This is demonstrated in table~2, where
 the resulting values  for $\fhat$ are given for all six
wave functions as a function of $t_{min}$,
the minimum time separation used in the fit to the local-smeared
correlator. The stability of $\fhat$, under variations of the wave
function and $t_{min}$, is remarkable. We therefore corroborate
our previous conclusion~\cite{Tsuk} that there is no
spurious dependence of $\fhat$ on the size of the wave function.

In addition, we have tried an alternative procedure, based on
the relation
\be
 C^{I,J;loc,loc}_{\gamma_4\gamma_5}(t) ~/~
\sqrt{C^{I,J;I,J}_{\gamma_4\gamma_5}(t) } \sim
<0|{\cal{M}}^{loc,loc}_{\gamma_4\gamma_5}(0)|P> \exp(-\tilde{M}_{P}~t/2)
\quad ,
\ee
which holds in the region where both correlation
functions on the {\rm l.h.s.} show ground
state dominance.
Performing  fits  to this ratio, we   obtain results
consistent with those from the  method described above.
But the
statistical errors are larger, because
one has to exclude part of the plateau of the smeared-smeared
correlator from the analysis\footnote{A
third possibility that has been discussed in ref. \cite{Allton}
differs only slightly from the method used here.}.

Let us finally mention that the optimal wave function ($\alpha=4,~n=100$)
carries an r.m.s. radius of about
$0.3 fm$, in agreement to earlier findings \cite{fb1}.

In completing our systematic study on the smaller lattices,
we restrict ourselves to the use of optimal wave functions, with
$r_2 \simeq 0.3 fm$.
In fig.~4 we show the t-dependent masses from the
local-smeared correlators for the
lattices A3, C3 and D2a corresponding to the $\beta$ values
 of 5.74, 6.0 and 6.26.
Note that the data points
again converge into their respective errors bands, as determined
from fits to the smeared-smeared correlators. The numerical
results for all the lattices are collected in table~3.
Here we quote values  on $\tilde{M}_{P}$ and $\fhat$ as well as indicate their
stability
with respect to different
 fit ranges in $t$, at various quark masses and $\beta$ values.
By looking at these numbers, we conclude that the results for $\fhat $
and $\tilde{M}_{P}$ are again
fairly insensitive to the wave functions chosen, once
the fit region is appropriately identified: a variation of the
lower cut $t_{min}$ in the fitting procedure renders results
for $\fhat$, that become stable, as soon as $t_{min}$ lies within the plateau.
In addition, we stress the agreement of the results for lattices A3 with A3a
and
C3 with C3a respectively. The old calculations~\cite{fb1} on A3a and C3a had
been done with the exponential wave function, whereas the results on A3 and C3
 are
obtained with $\Phi^G$. We find a beautiful stability of the results
with respect to  the $\it shape$ of the wave function. Furthermore,
$\Phi^E$ gives earlier plateaus at $\beta=5.74$ but at $\beta=6.0$ no
clear difference
is visible. This, together with the early plateaus achieved at $\beta=6.26$,
 indicate that $\Phi^G$ is the better choice at larger
$\beta$, a feature that is very welcome, given the low numerical cost
of constructing $\Phi^G$.

\subsection{Mass splittings}

We start with the determination of the mass difference
 between the scalar and pseudoscalar
states,
\be
\Delta_S = \tilde{M}_S-\tilde{M}_{P} = M_S-M_{P}  \quad .
\ee
For the scalar state we will discuss results that have been obtained with
the interpolating field ${\cal M}^{I,J}_{1}(\vec{x},t)$ from eq.(\ref{meson}).
In order to reduce statistical and systematic errors
the mass difference, in most cases, is determined directly from an analysis of
the ratio of
the respective correlators:
\be
R^{I,J}_{S}(t) =C^{I,J;I,J}_{1}(t)  ~/~ C^{I,J;I,J}_{\gamma_4\gamma_5}(t)
\label{ratios}
\ee

We find that the scalar ground state signal is somewhat inferior to
the respective pseudoscalar signal.
We note in passing that we have attempted  to improve signals
by using another operator to excite the scalar state:
In the non-relativistic quark model, the scalar state is
a p-wave excitation. We therefore tested in addition an interpolating field
${\cal M}^{P,loc}_{\gamma_4\gamma_5}(\vec{x},t)$ with a p-wave trial wave
 function
$\Phi^{P}$ which is constructed
 by applying a
covariant (lattice-) derivative to the wave functions from sect.~3.2.
The latter method produced consistent results,
but with larger statistical errors.

The t-dependent masses from smeared-smeared
correlators as derived from $R_S^{I,J}$ are plotted in fig.~5.
The x-axis is scaled to physical units using the string tension.
Comparing results obtained at different volumes, the figure demonstrates
that
finite size effects are negligible for $\Delta_S$.
We find first evidence of
plateaus in the regions $t\sqrt{\sigma}\geq 1$ for $\beta=5.74$
and $t\sqrt{\sigma}\geq 0.7$ for $\beta=6.0$ and $\beta=6.26$.
However due to the early
noise dominance, the situation is not as clean as in the case of the
pseudoscalar yielding to a larger uncertainty in the determination of
$\Delta_S$.
The numbers for $\Delta_S$, as obtained from a one-term exponential fit to
eq.(\ref{ratios}), are
quoted in table~4.
 In a few cases where the plateau for $R_S$ was very limited
we fitted separately the two channels to have at our disposal somewhat
larger plateaus at a price of larger error bars. These fits are indicated by
the daggers in table~4.

In an analogous way, we can determine the mass difference between
the $\Lambda_b$ and the $B$-meson
\be
\Delta_\Lambda = \tilde{M}_\Lambda-\tilde{M}_{P}= M_\Lambda-M_{P}.
\ee
The form of the
$\Lambda$-correlator reads
\be
D^{I,J}(t) = \sum_{\vx} < Tr \{ \gamma^+
{\cal B}^{I,J}(\vx,t)~ [ {\cal B}^{I,J
}(\vec{0},0)]^\dagger \}  > \quad ,
\ee
where the $\Lambda$-operator $\cal B$ is defined by
\be
{\cal B}^{I,J}_\alpha (\vx,t) = \sum_{a,b,c,\alpha,\beta,\gamma}
\epsilon_{abc} ~ (h^I (\vx,t))^a_\alpha ~
(u^J (\vx,t))^b_\beta ~ (C\gamma_5)_{\beta \gamma} ~
(d^J (\vx,t))^c_\gamma \quad .
\ee
Here a,b,c ($\alpha,\beta,\gamma$)
denote color (Dirac) indices and $C$ is the charge conjugation matrix.
The field ${\cal B}^{I,loc}_\alpha (\vx,t)$
is to be interpreted as
a diquark trial wave function for the baryon:  the two light quark fields
are
restricted to be local with respect to each other.
Otherwise ${\cal B}^{I,J}_\alpha (\vx,t)$
amounts to a general wave function ansatz. In ref.~\cite{Boch} where the mass
of $\Lambda_b$ was computed a diquark type of wave function was used.

Like in the case of $\Delta_S$, we avoid computing the mass splitting by
 separate
analysis of the individual correlators, but rather consider their ratio
\be
R^{I,J}_{\Lambda}(t) =
D^{I,J} (t)  ~/~ C^{I,J;I,J}_{\gamma_4 \gamma_5}(t) .
\label{Lambdarat}
\ee
We plot the t-dependent mass-splitting arising from $R^{I,J}_{\Lambda}(t)$
in fig.~6, where both axis are scaled to physical units using, as in
fig.~5, the string tension.
This means that all data should, apart from scaling violations,
merge into one and the
same plateau. The data are  consistent with  scaling,
but the statistical errors are obviously  too large to
make a definite case.
Note that the
t-dependent mass-splitting for $\beta=6.26$ where a diquark trial wave function
is used
starts off very high. The effect
is due to the numerator in eq.(\ref{Lambdarat}). Therefore
the underlying diquark type
trial wave function achieves a
small overlap with the ground state.
The other trial wave functions appear to do  better.

In table~4, we have included the results arising from one-term exponential fits
to  eq.(\ref{Lambdarat}). The fits were done in the t-ranges
where the t-dependent mass-splittings are consistent with a plateau.
We stress, however, that the existence of these plateaus is still debatable.
In order to settle this issue more effort must be spent to optimize
the wave functions.

\section{Connecting to Physics}

\subsection{The Renormalisation Problem}
We  have encountered various quantities in the static
approximation on the lattice, whose physical significance
has to be elucidated. The most questionable quantity
is the ``pseudoscalar mass'' $\tilde{M}_P$,
as it carries a linear divergence, due
to the self energy. In this section we will first investigate whether
lowest order effective coupling perturbation theory
allows to extract this linear divergence.
If this worked out, one would be able to produce  meaningful
numbers on the binding energy $B$ from static lattice calculations.
Another important issue is the renormalisation
factor $Z_{stat}$ for
the axial current. We are going to consider the possible improvement
on this perturbatively computed quantity,  by employing a renormalised
coupling.

{\bf Binding energy.}
We start by splitting  up the mass
into the binding energy $B$ and the divergent self energy $E(a)$:
\be
\tilde{M}_P = E(a)+B \quad .
\label{mp}
\ee
An estimate for the self energy may be obtained in perturbation theory,
with
the lowest non-trivial order term being  linearly divergent
\bes
E(a)|_{pert}=\frac{1}{a}( \frac{2}{3} G(\vec{0})
\twiggle{g}^2~+~O(\twiggle{g}^4)),\label{selfen}
\\
G(\vec{0})=
\int_{-\pi}^{\pi}\frac{d^3k}{(2\pi)^3}\frac{1}{4\sum_j
\sin^2(k_j/2)}=0.253(1) \quad .
\nonumber
\ees

The coupling in eq.~(\ref{selfen}) should be taken at   the scale
of the cutoff $a^{-1}$. It has been suggested~\cite{gimpr}
that a more suitable  expansion parameter than the bare coupling,
$g_0^2=6/\beta$, at this scale is
given by
\be
\twiggle{g}^2=g_0^2/<\frac{1}{3}\tr P_{\mu\nu}>\quad .
\label{gsq}
\ee
In addition one observes~\cite{LM}, that the $g^4$ terms of the perturbative
expansions of several short distance quantities are much smaller
in an expansion in terms of $\twiggle{g}^2$ as compared to an expansion
in terms of $g_0^2$, which points to a better convergence of the former
expansions. Finally, it has been shown that indeed the 1-loop
relation between $\twiggle{g}^2$ and $g^{2}_{\overline{MS}}$ is quite
accurate in SU(2) pure gauge theory~\cite{gr}. It should be noted, moreover,
that the use of an effective coupling similar to  eq.~(\ref{gsq}), in the
perturbative
expansion of the $\beta$-function, amounts to a considerable reduction of
subasymptotic contributions to the string tension~\cite{BS1}.
Therefore, to present knowledge, $\twiggle{g}^2$ is the appropriate coupling
to be used in the perturbative expansion eq.~(\ref{selfen}).

It remains to be checked, however, whether a perturbative estimate
of $E(a)$ is sufficient to extract the binding energy $B$. This can be done
by looking at the resulting estimate of the binding energy~\footnote
{It should be noted
that eqs.~(\ref{mp}) and (\ref{selfen}), are expected to hold at
most for an intermediate range of lattice spacings, since,
as one approaches the continuum limit $a \rightarrow 0;~\twiggle{g}^2
\rightarrow 0$, possible
non-perturbative
terms in eq.~(\ref{selfen}) become increasingly more important when
they are inserted into eq.~(\ref{btilde}).
}:
\bes
\tilde{B} \equiv \tilde{M}_{P} - E(a)_{pert} \quad .\label{btilde}
\ees
It should exhibit  scaling behavior, up to
terms of order $O(a)$. Using data from table~3 we obtain
for the dimensionless ratio
$\tilde{B}/\sqrt{\sigma}$  the values 1.40(2), 1.80(6), and  1.98(6)
 at $a\sqrt{\sigma}=$ 0.38, 0.22, and 0.151, respectively.
As the observed  variation of $\tilde{B}$
might still constitute  a large $O(a)$ effect, we have in addition
computed  the quantity
$$
(V_0-2~E(a)|_{pert})/\sqrt{\sigma} \quad ,
$$
with $V_0$ the constant piece of the heavy quark potential defined
in eq.~(\ref{pot}) and given in table~5.
It should also be independent of $a$ if $E(a)|_{pert}$ is a good
approximation to the divergent self energy of a static quark. In this case
the corrections should be smaller - of order $O(a^2)$.
Here, the numbers read
 $(V_0-2~E(a)|_{pert})/\sqrt{\sigma}=$ 0.008(16), 0.28(3),
and 0.56(2) to be compared
to $V_0/\sqrt{\sigma}=$ 1.70, 2.81, and 4.02,
 respectively.
We conclude that on this level
there is no justification whatsoever
to estimate  $E(a)$ by lowest  nontrivial order perturbation theory.

We have discussed this issue in some detail in view of the
recent proposal
to  simulate heavy quark systems with a nonrelativistic effective
Lagrangian~\cite{NRQCD}.
In practice~\cite{Davies,LepTsu}, the coupling constants in this effective
 Lagrangian are estimated
from perturbation theory to first order in $\twiggle{g}^2$.
They are divergent like $E(a)$. The  above considerations  cast
considerable doubt  on the validity of
such a procedure or at least call
for a serious estimate of systematic errors introduced through
the uncertainties in the coupling constants in the effective Lagrangian.

As a result, we have to refrain from quoting any number
for the binding energy in a static--light meson.
On the other hand,
in mass splittings like $\Delta_S$ and $\Delta_{\Lambda}$, the self
 energies
cancel and we obtain relevant estimates that are good to lowest order in
$1/m_h$.

{\bf Decay constant.}
We turn next  to the question of optimizing, within a given
order, the perturbative renormalisation factor $Z_{stat}$.
This quantity is implicitly defined \cite{BLP,BLPnew}
 by
matching the continuum (full theory) axial vector correlation function
$C_{A_0}(t)$
at large euclidean time
separation $t$  with its lattice
counterpart $C^{loc,loc;loc,loc}_{\gamma_4\gamma_5}(t)$:
\be
 C_{A_0}(t) e^{M_H t} = Z^2_{stat}(a,m_h) e^{\tilde{M}_P t}
 C^{loc,loc;loc,loc}_{\gamma_4\gamma_5}(t)~(1 + O(1/m_h))\quad .
\label{match}
\ee
Since eq.(\ref{match}) has to hold also in the range of intermediate
 time separation
$t$, the renormalisation constant $Z_{stat}$ can be calculated in perturbation
theory:
\be
  Z^2_{stat}(a,m_h) = e^{(M_H-\tilde{M}_P) t}
\frac{ C^{loc,loc;loc,loc}_{\gamma_4\gamma_5}(t)}
{C_{A_0}(t)} |_{pert} \quad .
\label{zstatdef}
\ee
 $M_H-\tilde{M}_P=m_h-E(a)$ has to be determined in perturbation theory only
to remove the $t-$dependence of the {\rm r.h.s.} and there is no
linear divergence in $Z_{stat}$~\cite{BLPnew}.

The result is:
\be
 Z_{stat}(a,m_h) = 1+ \frac{{g}^2}{4 \pi} (
\frac{1}{\pi} \log(a~m_h) - 2.372) \quad .
\label{zstat}
\ee
This is a first order perturbative result and it is not {\it a priori}
clear which coupling should be used in eq.~\ref{zstat}.
It is common practice  (see
ref.~\cite{BLPnew} and references therein) in numerical estimates of
 eq.(\ref{zstat})
and hence of $\fhat$ to use the bare coupling $g_0^2$  as  this choice seems
to be natural in connection with lattice regularisation. However, the
expansion in the coupling constant has to approximate a physical correlation
function both in the continuum (left hand size of eq.~(\ref{match})) and on
the lattice (right hand size of eq.~(\ref{match})). As we have already pointed
out in connection with eq.~(\ref{gsq})
$g_0^2$ is not a good expansion parameter and should be
replaced by a renormalised coupling constant defined by some physical
process~\cite{LM} .
We may use the $\overline{MS}$ coupling at a suitable scale which is known to
give well behaved expansions
for many physical quantities. For this purpose we have
to estimate the appropriate scale and establish the relationship between
$g_{\overline{MS}}^2$ and $g_0^2$.  To estimate the appropriate scale   in
eq.~(\ref{zstat}) we note that
 eq.~(\ref{zstatdef}) is valid for $t>>1/m_h$,
due to the expansion in $1/m_h$.
On the other hand,  the time separation is required
 to be
small compared to a typical non-perturbative scale,
such that the
 correlation functions can be approximated by perturbation theory.
This sets the requirement
$t<<0.5fm$.
For $m_h = O(m_b)$, these requirements don't leave much of a window and we
conclude that the coupling constant should be evaluated at a scale of about
3 GeV.
An estimate for this
coupling in pure SU(3) gauge theory is given by~\cite{LM}
\be
g^{-2}_{\overline{MS}}(\frac{\pi}{a}) = g^{-2}_{0} <\frac{1}{3}\tr P_{\mu \nu}>
 + 0.02461~~.
\label{gmsb}
\ee
This refers to the $\overline{MS}$ coupling in the continuum of
the pure SU(3) gauge theory.
 The corresponding expression
has been shown to be rather accurate at values of the lattice spacing
$a^{-1}=4$GeV in the pure SU(2) gauge theory~\cite{gr}.
As input into eq.(\ref{gmsb})
we need, in principle, an estimate for the lattice spacing at a small
value of the bare coupling constant.
Using the values for the string tension in table~5 (and for the larger
 $\beta$, the
2-loop renormalisation group equation) we obtain consistently
$$
g^{2}_{\overline{MS}}(3 GeV)=1.9(1) \quad ;
 \quad Z_{stat}(a,m_h) =0.71 + 0.048 (\log(am_h)-1.43) \quad .
$$
Let us estimate roughly the uncertainty of this renormalisation constant.
The logarithmic correction varies only by $\pm 0.007$
over  our range of
lattice spacings at $m_h=4.6 GeV \simeq m_b$.
Varying the argument of
$g^{2}_{\overline{MS}}$  between  1.5  $GeV$ or 9 $GeV$
produces  a maximal change in $Z_{stat}$ of $\pm 0.08$.
This range also covers changes arising from using a different quantity
like the $\rho$-mass to set the scale and a change to
other renormalised couplings like
$g^{2}_{MS}$ or a coupling defined from the force between heavy
quarks~\cite{BS1}
\footnote{Note, however, that this range does not include
the bare coupling constant.}.

\subsection{Setting the Scale}

We obtain physical results by forming dimensionless
ratios, determining their value through the simulation and identifying
one quantity with its experimentally measured value. In this way,
a reference scale enters. Within the quenched approximation, a
natural choice is the {\it string tension} $\sigma$, since it does not involve
an extrapolation in quark mass. It carries
the disadvantage, though,
 that it can only be determined from experiments in a rather
indirect way.

String tension calculations in lattice gauge theory
 mainly bear  two sources of systematic
errors. Firstly the individual values of the potential are determined at
{\rm finite} values of the euclidean time separation,
which makes them prone to pollution by
``excited states''. This
error is presumably
 smaller \cite{BS,MTC} than the statistical errors of the
measurements used here. Secondly the potential
parameters are determined
from fits within a certain range of the quark separation $R$.
In order to study
scaling violations effects, it is crucial to
use potential parameters, that have been
obtained from fits to  a definite parametrisation and fit range in physical
 units.
Here, we follow the philosophy of
employing the data from ref.~\cite{BS}
and \cite{MTC}, and restricting ourselves to
the range 0.3 $fm$ $\leq$ R $\leq$ 1.0 $fm$\footnote{We are aware of
the effects from
the subleading
terms and other overall uncertainties due to the
parametrisations used.
We expect that they have little impact on the
size of scaling violations, however.}.
In the following, we use results for $\sigma$
extracted from fits done with one
assumption on the subleading term, namely the one suggested by
the bosonic string picture~\cite{Lues}:
\be
a~V(\vec{R}) = a V_0 - {\pi^2\over 3} G(\vec{R}/a) + a ~ \sigma~R,~~
R=|\vec{R}|,~~
for ~R\sqrt{\sigma}>0.3
\label{pot}
\ee
In the $R$-range given in eq.~(\ref{pot}),
such a parametrisation of the potential
is in agreement with all existing lattice simulation
results.

Apart from the direct results obtained in simulations, table~5  contains
interpolated values for $\beta=$ 5.74 and 6.26. For these two $\beta$
values, $V_0 a$ and
$\log(\sigma a^2)$ have
been interpolated linearly in $\beta$ using the two
neighbouring Monte-Carlo results.
Some uncertainty due to possible deviations from such behavior has been
taken into account. Using the values collected in table~5
we can thus express all lattice quantities in physical units.

A second choice for setting the scale is
 the mass of the $\rho$ resonance $M_{\rho}$.
Before the $\rho$ mass can
be utilized to set the scale, its lattice prediction has to be
extrapolated into the chiral regime, however.
On the lattice, an accurate computation of
$M_{\rho}$ has been performed
in  the quenched approximation at $\beta=5.7$, $\beta=6.0$ and
$\beta=6.3$
by the APE group~\cite{APEmrho}. Our own determinations are listed in table~6:
$M_{\rho}a = 0.534(11)$, 0.341(15) and 0.208(15) at
$\beta=5.74$, 6.0 and 6.26 respectively. The $\beta=6.00$ value is in perfect
agreement  with the result of ref.~\cite{APEmrho}, although we have used
a significantly smaller lattice.

{\bf Extrapolation in Quark Mass.}
For completeness, we will now present some
details on quark mass extrapolation.
One makes a linear ansatz for the vector meson mass
\be
a~M_V(l,l') =B + C~ (\frac{1}{2\kappa}+\frac{1}{2\kappa'}-\frac{1}{\kappa_c})
\quad .
\label{mro}
\ee
The critical value  $\kappa_c$ follows from the  relation
\be
a^2~M^2_P(l,l') =A ~ (\frac{1}{2\kappa}+\frac{1}{2\kappa'}-\frac{1}{\kappa_c})
\label{pcac}
\ee
suggested by lowest order chiral perturbation theory~\cite{GaLe}. By
$l$ and $l'$ we denote quantities obtained using correlators of light quarks.
Here, $\kappa_c$ differs from its free value of 1/8 because of the
explicit breaking of the chiral symmetry in the lattice
regularisation~\cite{KaSm}. Our fits to eq.~(\ref{pcac}) are shown
in fig.~7, for degenerate and non-degenerate
light quark masses. The linear dependence of both $M_P^2(l,l')$ and
$M_V(l,l')$
on the quark masses is
very well satisfied.
The $\kappa$ values corresponding to the up and down quark ($\kappa_u$)
 and to the strange quark
($\kappa_s$) can be extracted from
$$
A ~ (\frac{1}{\kappa_u}-\frac{1}{\kappa_c}) / (\sigma a^2)=
(137MeV/420MeV)^2
$$
and
$$
A ~ (\frac{1}{2 \kappa_s}+ \frac{1}{2 \kappa_u}-\frac{1}{\kappa_c}) / (\sigma
 a^2)=
(494MeV/420MeV)^2
$$
when the string tension $\sigma a^2$ is used to set the scale
(with analogous expressions when scaling with  $M_{\rho} a$).
In addition, we quote results for
a large number of $L$ and $a$ values,
computed at a  quark mass $m_{2s}$ of approximately
twice the strange quark mass defined  by $M^2_P(l,l)/\sigma=4$.
They do not
require  any extrapolation, but only a mild interpolation in the case
of the $\beta=6.26$ data.

In table~6 we list our results for the vector meson mass  at
$\kappa_c$ and  $\kappa_s$.
Tables~7 and 8 give the values,
after linear extrapolations to
zero quark mass and to  $m_s$\footnote{The
values of these quantities
at $m_{2s}$ can be directly obtained from tables~3 and 4.},
for the bare decay constant and mass-splittings, respectively.

As we will base our analysis in the following on the scales
extracted from the string tension and the $\rho$-mass, we briefly
investigate, in how far we can expect consistent results from
these two scales. Figure 8 shows the ratio $m_\rho/\sqrt{\sigma}$
as a function of the lattice spacing. Perturbative arguments suggest
that the dependence of physical ratios on the lattice spacing is
linear in the case of the Wilson action\footnote{
 It should be remembered,
that this is a perturbative argument. It is
in no way obvious that this behavior is the same for
long-distance, nonperturbative quantities.}.
A linear extrapolation in the lattice spacing is shown in the figure.
The result at $a=0$ is some 10\% low from the experimental value
of 1.75-1.83.

Let us comment finally on the possibility to base the
scale on the $\pi$ decay constant.
Our data for $f_{\pi}$ show
 non-linear behavior in the quark mass.
 Recent calculations of $f_{\pi}$ have been able to
reach smaller quark masses with good precision and can therefore be
extrapolated~\cite{QCDPAX}.
 Within errors,  $f_{\pi} a$ and $M_{\rho} a$ give
the same lattice spacing  at $\beta=5.8$ and 6.0.
The situation for our data is demonstrated in fig.~9.
Taking into account the uncertainty in the perturbative renormalisation
of $f_P$, which is not included in fig.~9, we see that the Monte Carlo
results for $f_P/M_V$ are
consistent with the experimental numbers and therefore the lattice spacings
agree within errors.
We therefore use
exclusively  the string tension and the  $\rho$ mass to convert
our results to physical units.

\subsection{Finite Size Scaling}

In fig.~10 we plot $\fhat/Z_{stat}$ as a
function of the spatial lattice
length L, both  in units of the string tension, at quark mass $m_{2s}$
twice the strange quark mass.
We wish to make two comments on the finite
size behavior of these data:\\
1.~ As the entries into fig.~10 are statistically independent,
an alternative search for possible $a-$effects
may proceed via
a global fit to the $L$-dependence. The latter
is exponential, with a characteristic decay given by the mass of
lowest lying glueball, which winds around the
periodic volume.
Since $m_{0^{++}} / \sqrt{\sigma} \simeq 3.5$ \cite{MiTe},
the sensitivity of our computer experiment is much too
low to exploit this.

2.~ One might speculate  that  the finite size effects
seen within the intermediate
regime of fig.~10 originate from a distortion of the
light-quark wave function in the finite volume.
For an exponentially decaying wave function,
the finite size effects on the wave function at the
origin are again exponential in $L$.
The characteristic scale in this domain is the coefficient in the
falloff of the wave function. A rough estimate is given by 1.5 in units of the
string tension~\cite{BeTall}.
We find that the data in fig.~10 can indeed  be fitted
with the form  $\fhat \sigma^{-3/4}=
3.64(8) - C \exp (-1.5 \sqrt{\sigma} L)$ with a $\chi^2/DoF=8/9$, if the
error of the most precise point is increased to lie in the general
ballpark of $\pm 0.15$.

{}From the above observations we may conclude that
within a precision of about 10 percent one can neglect
finite size effects on $\fhat \sigma^{-3/4}$, in the range
$\sqrt{\sigma} L \geq 3$.
This remains true also for smaller quark masses, as can be seen
by looking at the entries in table~3 for the lattice A3, A3a and A5.
We therefore assume that finite size effects are negligible
beyond $\sqrt{\sigma} L \geq 3$ for the light quark masses considered here
and we proceed to analyse our results in more detail
at $\sqrt{\sigma} L \simeq 3$.

Fig.~11 shows the $a$-dependence of the bare decay constants at the
s quark mass and at the u quark mass with both setting the scale
through $\sqrt{\sigma}$ and $M_\rho$.
Putting in the value $Z_{stat} = 0.71(8)$ we obtain
\be
\fhat_s=0.61(4)(7)GeV^{3/2},~~~\fhat_u=0.53(5)(6)GeV^{3/2} \label{fsig}
\ee
with $\sqrt{\sigma}$ and
\be
\fhat_s=0.67(18)(7)GeV^{3/2},~~~\fhat_u=0.59(18)(6)GeV^{3/2} \label{frho}
\ee
from the $M_\rho$-scale.
The values given in eq.~\ref{fsig} and \ref{frho} are in perfect agreement,
with the latter ones
having very large errors. The first errors given above are statistical
and the second result from our estimated uncertainty
of $Z_{stat}$. We have not included an error for the extrapolation, since
our data points at the smallest value of $a$ are within
the error bar of the extrapolated values.

{}From eq.~\ref{fsig} we obtain the IMET lattice predictions for the decay
constants of B-mesons evaluated in the {\it static approximation}

\be
f_{bs}=266 \pm 22 \pm 26~~ MeV \hspace*{3cm} f_{bu}=230 \pm 22 \pm 26~~ MeV.
\ee

The central value is significantly lower than the ones of the first
calculations \cite{fb1,Allton}. Firstly, there is a significant a-dependence
when the scale is set by the $\rho$-mass, as anticipated in ref.
\cite{fb2}. Secondly, as we explained in section~4.1 the use of
{\it renormalised} perturbation theory in the determination of $Z_{stat}$
decreases the value of the decay constant by about 13\%.

For the D-meson one definitely needs the $1/M_P$ correction since by
stretching the validity of IMET down to the charm mass, we obtain the
extremely high values of $f_{cu} = 359(34)(39)$ and
$f_{cs} = 405(27)(46)$ MeV which contradict
results both from lattice calculations with   fully propagating quarks and
from  sum rule evaluations~\cite{SumR}.

If one looks at the analogue of fig.~10 for the 1P - 1S splitting,
$\Delta_S$, one observes a significant
scatter in the data coming from different $\beta$'s. This
indicates that $\Delta_S/\sqrt{\sigma}$ has a stronger $a$-dependence than the
decay constant. Looking at the results given in Table~4, however, we find
no significant finite size
effects for $L\sqrt{\sigma} \gae 2$ on the level of our statistical
uncertainties.
Therefore, we can extract the splitting directly at $L\sqrt{\sigma}\simeq 3$.
In fig.~12, we show the $a$-dependence of
$\Delta_S$ at quark mass $m_{2s}$.
The linear extrapolation to the
continuum gives 344(37) MeV from fig.~12a and 383(72) MeV from fig.~12b.
At smaller quark masses, our results are too inaccurate to allow for such
an extrapolation. We note, however, that no significant quark-mass
dependence was observed for this splitting.

Unfortunately a similar analysis is not feasible at the moment for
the Baryon-Meson splitting, $\Delta_\Lambda$.  The best we can do, is
to take figure 6 as an indication that there is a plateau at
$\Delta_\Lambda/\sqrt{\sigma}\simeq 1.5$.  This produces  a rough estimate
of 600MeV for the $\Lambda_b$ -- B-meson mass splitting.  Much more
work is needed to check the existence of plateaus, the $a$-dependence,
quark-mass dependence and finite size effects for this quantity.

\section{String Breaking}
We turn now to another observable that may be calculated from our simulation
results.
It is of interest, because it gives
information about the forces between static
charges in the {\it full} theory, i.e. it is a quantity
calculated in the quenched approximation which can provide
information about $q\bar{q}$ forces with dynamical fermions.
In full QCD simulations
 the breaking of the
QCD string, i.e. the flattening of the heavy quark potential at large
distances,
has been searched for some time.
No effect was found in the most serious effort~\cite{MTCTsu}. In the following,
we   estimate
the distance $R_b$, where the full QCD potential
flattens off using only quantities calculated in the quenched
approximation~\cite{Tsuk}.

Consider a large Wilson loop $W({\vec R},T)$,  with~~$ T>>R$, in full QCD.
It has a representation in terms of the eigenvalues of the QCD-hamiltonian
(or transfer matrix):
\be
W({\vec R},t) = \sum_{n \geq 0}
 |c^W_n({\vec R})|^2
             \exp( - V_n({\vec R}) t) \quad . \label{Wloop}
\ee
Here $\exp( - V_n({\vec R}))$ are the eigenvalues of the transfer matrix in the
corresponding charged sector of the Hilbert space:
the states in this sector transform under gauge transformations according
to the 3-representation at  position $\vec{0}$ and according
to the $\bar{3}$-representation at  position ${\vec R}$.
 The same states contribute in the spectral
 decomposition of
the
correlation function
\bes
H({\vec R},t) &=& <  {\cal M}^{loc,J}_{\gamma_4\gamma_5}({\vec 0},t)~
[ {\cal M}^{loc,J}_{\gamma_4\gamma_5}({\vec R},t)]^\dagger
{\cal M}^{loc,J}_{\gamma_4\gamma_5}({\vec R},0)~
[ {\cal M}^{loc,J}_{\gamma_4\gamma_5}({\vec 0},0)]^\dagger    >  \nonumber \\
&=&
\sum_{n \geq 0}
 |c^H_n({\vec R})|^2
             \exp( - V_n({\vec R}) t ) \quad .    \label{H}
\ees
The ground state potential $V({\vec R})=V_0({\vec R})$ can therefore
be called a static quark
potential or a static meson potential. Physically, the
first interpretation is sensible at relatively short distances, whereas
the second one is the appropriate language for large distances (compared
to the confinement scale). This can even be put into a quantitative
relation: we expect $ |c^H_0({\vec R})|^2 << |c^W_0({\vec R})|^2$ at
 small distances and $ |c^H_0({\vec R})|^2 >> |c^W_0({\vec R})|^2$ at large
distances. Furthermore, at large distances, the potential will approach
a constant up to non-leading terms (Yukawa-type interactions), because the
correlation function factorises:
\be
log[H({\vec R},t)] = log[( C_{\gamma_4\gamma_5}^{loc,J;loc,J} )^2]
+ O( \exp[ - m_{\pi}R] t/R )\quad .    \label{factor}
\ee
Simulation results of full QCD~\cite{MTCTsu}
 indicate, that the QCD-potential is
approximated rather well by the quenched potential out to relative
large distances,
up to about $R \sim 0.7$$fm$.
At very large distances, on the other hand, we expect
that $H({\vec R},t)$
is represented with some accuracy
by the quenched approximation. We must
switch from one correlation function to the other when using the
quenched approximation, since we have to put in the breaking of the
string by hand. Obviously, the quenched approximation does not have
much to say about the intermediate regime.

The asymptotic behavior of the correlation function $H({\vec R},t)$
 is given by the
mass in the static approximation.
So we define the string breaking distance  $R_b$ by
\be
V(R_b) = 2 \tilde{M}_{P},
\label{Rb}
\ee
with $V(R)$ being the quenched potential including the self-energy term that
cancels in eq.(\ref{Rb}).
$R_b$ defined in this way, gives an upper bound
to the distance where the potential starts deviating significantly from
the form of eq.(\ref{pot}).

{}From eq.(\ref{pot})
 with $G({\vec R}/a)\simeq a/(4\pi R)$,
 we can calculate $R_b$
\be
R_b=(\tilde{M}_P-\frac{1}{2} V_0)/\sigma + \sqrt{[(\tilde{M}_P-\frac{1}{2}
V_0)/\sigma]^2+
\frac{\pi}{12\sigma} }
\ee
As table 3 does not show any significant finite size effects on $\tilde{M}_P$
within
the whole range of $1.5 \leq L \sqrt{\sigma}\leq 4.5$, we can safely
neglect finite size effects on $R_b$.
We have extrapolated $\tilde{M}_P$ at $L \sqrt{\sigma}\simeq 3$
linearly in $1/\kappa$ and calculated $R_b$
at $\kappa_s$ and $\kappa_u$. The results are listed in Table 8.
Figure 13 shows the $a$ dependence and our extrapolation to $a=0$.
For all quark masses, the continuum value is at $R_b=4/\sqrt{\sigma}$
with about 10\% statistical error. We estimate an uncertainty for
the extrapolation in $a$ of 10\%, since the $a$-dependence is rather steep
and end up with
\be
R_b=1.9(2)(2) fm \quad
\ee
for all quark masses $m_q < 2 m_s$.

Such distances are difficult to reach in a potential calculation
including dynamical fermions.
It should be noted, however, that the screening of the potential is expected
to appear at distances that are rather independent of the
dynamical quark mass and hence can be observed with relatively large
quark masses
\footnote{Note that the significant change compared to the numbers given
in ref. \cite{Tall}, is due to  a copying mistake in that reference.}.

\section{Summary and Conclusions}
In this work we have  investigated in detail the  various aspects of
the static approximation. We summarize here the main outcome of this study:
\begin{enumerate}
\item In order to filter out the ground state before noise sets in, one
must use extended fermion fields. By varying the trial wave functions
at $\beta=6.26$ we found that extended plateaus in the local masses
occur for wave functions with r.m.s. radius of about 0.3 fm.
Using wave function with larger or smaller radii, the quality of the
plateaus deteriorates.  In particular for the local-smeared correlators the
quality
of the plateaus depends crucially on the choice of a good trial wave
function. For the very small or large wave functions the plateaus set
in at larger times where errors become significant. For these wave
functions we can safely identify the region of the plateau only by
using our knowledge of the height of the plateau deduced from looking
at the smeared-smeared correlator. If we then fit in the range of the
identified plateau we arrive at results which are all consistent and
stable under variation of the time interval of the fit. The error bars
are smallest for the best wave function. The results are insensitive to
a change of shape of the trial wave functions of given r.m.s. radius.
We therefore conclude that the results for the decay constant in the
static approximation are stable under variations of the size and shape
of the trial wave functions as long as plateaus are correctly
identified in the manner discussed in section~3. This corroborates our
previous findings \cite{fb2}.
\item Having established the wave function independence of the decay constant
we studied scaling, using the best wave function at each value of $\beta$.
We find that for $L
\gae 1.5$~fm the volume dependence of $f$ is negligible. Using results
at a fixed volume of about $(1.5~{\rm fm})^3$ we observe the
a-dependence to be rather strong when the mass of the $\rho$ meson is
used to set the scale. If, on the other hand, the physical scale is
based on the string tension the a-dependence turns out to be
much weaker. In other words, at small $\beta$ values one is faced with
significant effects from the choice of scale.  After continuum
extrapolation, however, both alternatives are observed to yield
identical results within error bars. But the latter are considerably
reduced in the string tension option.
\item In our original
calculation of $f$~\cite{fb1} we anticipated systematic errors arising
from the renormalisation constant $Z_{stat}$, the choice of
physical scale and the $a \rightarrow 0$ extrapolation.  Here we
have carried out the $a \rightarrow 0$ extrapolation and found a 10\%
correction due to the scale and a 13\% correction due to the
renormalisation constant $Z_{stat}$. All these corrections act in the
same direction, i.e. they lower the estimate for $f$. As a
consequence, we finally end up with \footnote{After completion of this work
we became aware of the most recent (preliminary) data from the Fermilab group,
presented
by Eichten at the Amsterdam {\it Lattice '92} conference.
They are in agreement with our data
at the largest value of the light quark mass. However they appear
to differ by $25 \%$ after linear extrapolation to the chiral point.
We are unable, at this point, to
give a conclusive explanation for the discrepancy.}
 $$ f_{bd}=230(22)(26) MeV,
\hspace*{3cm} f_{bs}=266(17)(30) MeV.  $$
\item In  the mass range of the D-meson there is a substantial
 difference between
the result for $f$ obtained in the static approximation and using propagating
heavy quarks, which points  to large $1/M_P$ corrections.
\item The plateau of the local mass for the scalar correlator is not
as wide as for the pseudoscalar. This means that the result for the
scalar-pseudoscalar splitting $\Delta_S$ is less conclusive than
for the pseudoscalar mass. We do not observe a definite  dependence on the
light quark mass. Extrapolating to $a=0$ our results at light quark mass
twice the strange quark we estimate

$$ \Delta_S=344(37) MeV.$$

For $\Lambda_b$ the situation is even less clear and more study is
required to establish a reliable estimate.
\item From our quenched calculation we are able to produce  an upper bound
for the distance where string breaking should occur in the full QCD.
After extrapolating to $a=0$ we obtain

$$  R_b=1.9(2)(2) fm.$$

We do not observe a strong dependence on the light quark when we vary from
the chiral limit to quarks of mass twice the strange quark mass.
\end{enumerate}


\noindent
{\bf Acknowledgements:}
We have benefited from discussions with O. Pene and E. Eichten whom we also
thank for communicating his new (preliminary) results to us.

These calculations were carried out  on the NEC-SX3 at
the Supercomputing Center in
Manno, Switzerland and on the CRAY-Y-MP in J\"ulich, Germany. We thank
both institutions for their generous support.

\eject
\section{Appendix A: Simulation Parameters}
We work in the quenched approximation, using the standard Wilson
formulation \cite{wilsong,wilsonf} of the lattice action
\be
S = S_G + S_F
\ee
with
\be
S_{G}(U) = \beta \sum_x \sum_{\mu > \nu} \{
           1 - \frac{1}{6}{\rm Re}\; {\rm Tr}\; P_{\mu\nu}(x) \}
\ee
and
\bes
S_{F}(\bar{\Psi},\Psi,U) =
  \sum_x \{ \frac{1}{2\kappa}\bar{\Psi}(x)\Psi(x) -  \frac{1}{2}\sum_{\mu}
[
 &\bar{\Psi}(x)  (1-\gamma_{\mu})U_{\mu}(x)
  \Psi(x+a\hat{\mu})&                  \label{fermact} \\
 + & \bar{\Psi}(x) (1+\gamma_{\mu})
  U^{\dagger}_{\mu}(x)\Psi(x-a\hat{\mu}) &
 ] \}  \nonumber \\
 = \bar{\Psi}M \Psi\qquad . \qquad \qquad \; \;  &&  \nonumber
\ees
Here, as in most of the following, we have suppressed
 Dirac and color indices. $a$ denotes the lattice constant and
$P_{\mu\nu}(x)$ is the product of gauge parallel transporters $U_\mu(x)
\in $ SU(3) around an elementary plaquette and is given by
\be
P_{\mu\nu}(x) = U_{\mu}(x)U_{\nu}(x+a\hat{\mu})
                U^{\dagger}_{\mu}(x+a\hat{\nu})
                U^{\dagger}_{\nu}(x) \quad .
\ee
The gauge fields are periodic functions of $x$, and the fermion fields $\Psi$
are taken periodic in space over a length $L$ and antiperiodic in time
over a time extent $L_t$.
The hopping parameter $\kappa$ is related to the bare quark mass $m$ by
\be
\kappa = \frac{1}{2(ma+4)} \quad .
\ee
In order to quantify finite $a$ and finite size effects,
the calculations have been performed on a series of lattices which
differ in the coupling constant $\beta$ and in the lattice size $L$.
In table~1 we give the parameters and nomenclature for the various
lattices. The physical lattice spacing and size of the lattice,
obtained by fixing the string tension $\sigma$ to the value $(420 \MeV)^2$,
are given in the last two columns.

Our Monte Carlo procedure employs a hybrid over-relaxation
algorithm~\cite{over}
to produce independent gauge configurations. In this algorithm $N$ almost
microcanonical over-relaxation sweeps \cite{micro}
through the lattice are followed by one
standard Metropolis sweep\cite{metro}. This combination is called one
iteration. We separated our measurements by
100 iterations with N=1 at $\beta=5.74$, 200 iterations with N=2 at
$\beta=6.0$ and 150 iterations with N=3 at $\beta=6.26$.

As demonstrated in section 3.1, in order to obtain a reliable signal
for the correlation functions and thus for the physical quantities of
interest, we must smear the hadronic fields.
This requires calculating the product of a finite-mass quark propagator $S$
with
smearing wave function $\Phi(x,x_4)$. This product is given by
the solution of
\be
\sum_{y} M(x,y)[\sum_{z}S(y,z) \Phi(z,x_4)]=  \Phi(x,x_4)\qquad ,
\label{qprop}
\ee
where $M$ is the fermion matrix of eq.(\ref{fermact}).
The linear equation
(\ref{qprop}) is preconditioned through an even-odd partitioning of
the matrix \cite{evod} and then solved by the minimal residual
algorithm~\cite{minres}. In order to speed up the process further,
the solution at the respectively higher
quark mass is used as starting guess in the algorithm.
For the quark masses used in
this calculation, this combination saves a factor of 2-4 in cpu time compared
to the standard conjugate gradient method.
The accuracy needed in the matrix inversion is determined
by testing the convergence of {\it all} correlation functions that are used
in the calculation.

Let us briefly comment on the error analysis of the simulation results.
We have generally binned our results into groups of up to
5 measurements, keeping, however, a minimum of around $N_{bin}=20$ bins.
As to be expected, no significant bin-size dependence of the statistical
errors was found for the simulation parameters
given above.
The statistical error of a function $F(P^1,...,P^r)$ of quantities $P^i$
that are simple averages over configurations, was calculated using
\be
\Delta({F}) = \frac{1}{\sqrt{N_{bin}}} ~\sqrt{ \sum_{i,j=1}^{r} \frac{dF}{dP^i}
 \; {\rm cov}(P^i,P^j) \; \frac{dF}{dP^j}} \quad,
\ee
with the covariance matrix given by
\be
{\rm cov}(P^i,P^j)= \frac{1}{N_{bin}}  \sum_{k=1}^{N_{bin}}
(P_k^i - \bar{P^i})(P_k^j - \bar{P^j}) \quad
 .
\ee
$P_k^i$ is the average of $P^i$ in the $k^{\rm th}$ bin.
This procedure was applied e.g. for the time-dependent masses discussed in
section 3.1.
Statistical errors of parameters originating from fits to correlation functions
or ratios of correlation functions  as well as from fits extrapolating
in mass were always determined by the use of the jackknife
method~\cite{jacknife} (with binning to control autocorrelations).

\section{Appendix B: Variance of ``local-smeared" correlators}

In the following, we
consider the variance of the correlations $C^{l,l;I,l}_\Gamma (t)$ and
$C^{l,l;l,I}_\Gamma (t)$. Under two assumptions we derive an inequality
which corresponds to the behavior of fig. 2.

As we calculate the real part of the correlation functions,
there are two contributions (one from the operator squared, one from the
operator times its adjoint). Since the discussion is completely identical
for the two terms we write only the first one in the following.
In the case where we smear the light quark
on the ``source side" this is given by
\bes
\sigma^2_1\{C^{l,l;l,I}_\Gamma (t)\} &=&  \\
&&\sum_{\vx, \vy} <  {\cal M}^{l,l}_{\Gamma,1}(\vx,t)~ [
     {\cal M}^{l,I}_{\Gamma,1}(\v0,0)]^\dagger
{\cal M}^{l,l}_{\Gamma,2}(\vy,t)~ [
     {\cal M}^{l,I}_{\Gamma,2}(\v0,0)]^\dagger >
- [C^{l,l;l,I}_\Gamma (t)]^2 \nonumber\\
&=&  <  {\cal M}^{l,l}_{\Gamma,1}(\v0,t)~ [
     {\cal M}^{l,I}_{\Gamma,1}(\v0,0)]^\dagger
{\cal M}^{l,l}_{\Gamma,2}(\v0,t)~ [
     {\cal M}^{l,I}_{\Gamma,2}(\v0,0)]^\dagger >
- [C^{l,l;l,I}_\Gamma (t)]^2~~, \nonumber \label{B1}
\ees
Here we have added a flavor index $1,2$ to the quark fields: $
{\cal M}^{I,J}_{\Gamma,i}(\vx,t) = \bar{h}^I_i (\vx,t)
{}~\Gamma ~l^J_i(\vx,t)$. It keeps track of the proper contractions of
quark fields in eq.~(\ref{B1})\footnote{
This is necessary since we want to
calculate the variance of the observable that is defined {\it after} we
have
integrated out the fermion fields.}.
It can be shown that
\be
\sigma^2\{C^{l,l;l,I}_\Gamma (t) \}=
\sigma^2\{C^{I,l;l,l}_\Gamma (t) \}=
\sigma^2\{C^{l,I;l,l}_\Gamma (t) \}.
\ee
We proceed by expressing the first part of the variance through the
eigenstates of the transfer matrix. States $|B_1(\v0)B_2(\v0)>$
with two static mesons (with different flavors)  at $\vx=0$ contribute.
Denoting the energy of these states by $\exp(-E_n^{BB}(\v0) )$, we obtain
\bes
\sigma^2_1\{C^{l,l;l,I}_\Gamma (t) \} &+& [C^{l,l;l,I}_\Gamma (t)]^2=
\sum_{\vx, \vy} \sum_n \exp(-E_n^{BB}(\v0)~t) ~\gamma(\vy,\vx)~~, \\
\gamma(\vy,\vx) &=&\gamma(\vx,\vy) =
<0|{\cal M}^{l,l}_{\Gamma,1}(\v0,t)~
{\cal M}^{l,l}_{\Gamma,2}(\v0,t) |n;B_1(\v0)B_2(\v0)> \\
&&<n;B_1(\v0)B_2(\v0)|
\bar{l}_1(\vy)F^I(\vy,\v0;{\cal U}(0))~\Gamma~h_1(\v0) ~~
\bar{l}_2(\vx)F^I(\vx,\v0;{\cal U}(0))~\Gamma~h_2(\v0) |0> \nonumber
\ees
Correspondingly we have
\bes
\sigma^2_1\{C^{l,l;I,l}_\Gamma (t)\} &+&  [C^{l,l;I,l}_\Gamma (t)]^2 =\\
&&\sum_{\vx, \vy} <  {\cal M}^{l,l}_{\Gamma,1}(\vx,t)~ [
     {\cal M}^{I,l}_{\Gamma,1}(\v0,0)]^\dagger
{\cal M}^{l,l}_{\Gamma,2}(\vy,t)~ [
     {\cal M}^{I,l}_{\Gamma,2}(\v0,0)]^\dagger >
  \nonumber\\
&=& \sum_{\vx,\vy} <  {\cal M}^{l,l}_{\Gamma,1}(\vx,t)~
\bar{l}_1(\v0,0)F^I(\v0,\vx;{\cal U}(0))~\Gamma~h_1(\vx,0) \nonumber \\
&&{\cal M}^{l,l}_{\Gamma,2}(\vy,t)~
\bar{l}_2(\v0,0)F^I(\v0,\vy;{\cal U}(0))~\Gamma~h_2(\vy,0)      >
{}~~, \nonumber\\
\sigma^2_1\{C^{l,l;I,l}_\Gamma (t) \}&+&  [C^{l,l;l,I}_\Gamma (t)]^2 =
\sum_{\vx, \vy} \sum_n \exp(-E_n^{BB}(\vx-\vy) ~t) ~\alpha(\vy,\vx)~~, \\
\alpha(\vy,\vx) &=&\alpha(\vx,\vy) ~=~
<0|{\cal M}^{l,l}_{\Gamma,1}(\vx)~
{\cal M}^{l,l}_{\Gamma,2}(\vy) |n;B_1(\vx)B_2(\vy)> \\
&&<n;B_1(\vx)B_2(\vy)|
\bar{l}_1(\v0)F^I(\v0,\vx;{\cal U})~\Gamma~h_1(\vx) ~~
\bar{l}_2(\v0)F^I(\v0,\vy;{\cal U})~\Gamma~h_2(\vy) |0> \nonumber
\ees
For large values of $t$, the expressions for $\sigma^2$ are dominated by the
$n=1$ contribution. Let us assume that $E_1^{BB}(\vx)$ increases with
$\vx^2$, i.e. that the   interaction
between the static mesons is attractive.
Then $\exp(-E_1^{BB}(\vx-\vy) ~t) \simeq
\delta_{\vx \vy}~\exp(-E_1^{BB}(\v0) ~t)$ in the limit of large $t$.
Since $\alpha(\vx,\vx) =\gamma(\vx,\vx)$ we obtain
(in the limit of large $t$)
\be
\sigma^2_1\{C^{l,l;l,I}_\Gamma (t) \}-
\sigma^2_1\{C^{l,l;I,l}_\Gamma (t) \}\simeq
 \exp(-E_1^{BB}(\v0)~t) ~\sum_{\vx \neq \vy} \gamma(\vy,\vx)~.
\ee
Since we are using wave functions without nodes, $\gamma(\vy,\vx)$ is
expected to be positive for all $\vx, \vy$.
The second contribution to the variance is identical to the above
except that we have $B_1$--$\bar{B}_2$ correlation functions instead of
$B_1$--${B}_2$.

Therefore -- accepting the above assumption -- we arrive at the
conclusion that (at large $t$) the variance of the correlation function
 can be reduced
if we smear the heavy quark on the ``source side" compared to the other cases.
 In fig. 2,  this effect is there for all values of $t$, which is
expected following the argument given above if one assumes that
$E_n^{BB}(\vx)$ increases with $\vx^2$ for general n.

\newpage
\noindent
{\bf Table Captions \hfill}

\begin{enumerate}
\small

\item[\bf Table 1:]
Parameters of the lattices used for this work. The inverse lattice spacing and
the spatial extension
$L$ in physical units are obtained from fixing the string tension
to $(420MeV)^2$.
\begin{center}
\begin{tabular}{|c|c|c|c|c|c|}
\hline
lattice & $\beta$ &  $L/a,~L_t/a$ & no. config's. & $a^{-1}$/GeV & $L/fm$   \\
\hline
A1  & 5.74 &  4, ~24 & 404 & 1.12& 0.70 \\
A2  & 5.74 &  6, ~24 & 131 &  & 1.06 \\
A3  & 5.74 &  8, ~24 & 170 &  &  1.41\\
A3a  & 5.74 &  8, ~24 & 100 &  &  1.41\\
A4  & 5.74 &  10, ~24 & 213 &  &  1.76\\
A5  & 5.74 &  12, ~24 & 140 &  &  2.12\\
B1  & 5.82 &  6, ~28 & 100 & 1.32 & 0.90 \\
C1  & 6.00 &  6, ~36 & 227 & 1.88 & 0.63 \\
C2  & 6.00 &  8, ~36 & 100 &  &  0.84\\
C3  & 6.00 & 12, ~36 & 204 &  &  1.26\\
C3a  & 6.00 & 12, ~36 & 100 &  &  1.26\\
C4  & 6.00 & 18, ~36 & 27 &  &  1.89\\
D1  & 6.26 & 12, ~48 & 103 & 2.78 & 0.85 \\
D2  & 6.26 & 18, ~48 & 33 &  &  1.28\\
D2a  & 6.26 & 18, ~48 &43  &  & 1.28\\
\hline
\end{tabular}
\end{center}
\newpage
\item[\bf Table 2:]
Bare decay constant and mass  in lattice units,
using the static approximation for the $18^3 \times 48$ lattice at $\beta=6.26$
and
$\kappa=0.1492$. The smallest time separation in the correlators
 used for the fit is given by
$t_{min}/a$ and the largest is fixed at $t_{max}/a=17$.
An asterisk $(*)$ indicates that the fit was not accepted due to either
the missing of a plateau in the local masses in the region of
the fit or a value of $\chi^2$ larger than the one to be expected
including the correlations of the data.
\begin{center}
\vspace*{-0.7cm}
\pagestyle{empty}
\begin{tabular}{|c|c|c|c|c|c|}
\hline
$\alpha$ ~~$n$ & $r_1/a$ & $r_2/a$ &$a\tilde{M}_{P}$& $t_{\rm min}/a$ &
 $a^{3/2}\fhat/Z_{stat}$\\
\hline
loc+gaus &2.5 &  3.9&0.569(7) & 2  & $ 0.235(19)^{*}$ \\
         & &        &        & 4 & $ 0.223(18)^*$\\
         & &        &        & 6 &  $0.213(18)^*$\\
         & &        &        & 8 &  0.208(20)\\
         & &        &        & 10 &  0.205(21)\\
         & &        &        & 12 &  0.203(24)\\
\hline
loc+gaus & 3.5& 4.6 &0.566(7) & 2 & $ 0.221(19)^{*}$ \\
         & &        &  & 4 & $ 0.211(19)^*$\\
         & &        &  & 6 & $0.204(20)^*$\\
         & &        &  & 8 &  0.200(21)\\
         & &        &  & 10 &  0.198(22)\\
         & &        &  & 12 &  0.195(25)\\
\hline
loc+gaus & 4.1& 5.0&0.566(7) & 2 & $ 0.216(19)^{*}$ \\
         & &       &   & 4 & $ 0.208(19)^*$\\
         & &       &   & 6 &  $0.201(20)^*$\\
         & &       &   & 8 &  0.198(20)\\
         & &       &   & 10 &  0.196(22)\\
         & &       &   & 12 &  0.192(25)\\
\hline
 4 70  & 4.2 &4.6 & 0.566(7) & 2 &$ 0.201(15)^{*}$ \\
         & &      &    & 4 & $ 0.200(15)^*$\\
         & &      &    & 6 &  0.198(18)\\
         & &      &    & 8 &  0.195(18)\\
         & &      &    & 10 &  0.194(20)\\
         & &      &    & 12 &  0.190(22)\\
\hline
 4 100  & 5.2 & 5.6&0.565(8) & 2 & $ 0.188(15)^{*}$ \\
         & &          &      & 4 & $ 0.190(16)^*$\\
         & &          &      &6 &  0.190(17)\\
         & &          &      &8 &  0.189(20)\\
         & &          &      &10 &  0.188(21)\\
         & &          &      &12 &  0.185(25)\\
\hline
 5 160  & 6.1 & 6.5 &0.566(8) & 2 & $ 0.170(13)^{*}$ \\
         & &        &          & 4 & $ 0.176(15)^*$\\
         & &        &          & 6 &  0.180(18)\\
         & &          &        &8 &  0.181(20)\\
         & &          &        &10 &  0.181(21)\\
         & &          &        & 12 &  0.180(24)\\
\hline
\end{tabular}
\end{center}
\pagestyle{plain}
\newpage
\item[\bf Table 3:]
Bare decay constant and mass  in lattice units,
using the static approximation.
An asterisk $(*)$ indicates that the fit was not accepted due to a large
$\chi^2$  or the missing of a plateau in the local masses in the region of
the fit. A dagger $(\dagger)$ indicates that smearing was applied to both
heavy and light quarks.
\pagestyle{empty}
\begin{center}
{\hspace*{-2.5 cm}
\begin{tabular}{|c|c|c|c|c|ccccc|c|}
\hline
latt. & $\kappa$  & $K$ &  $\alpha$ ~~$n$ & $r_2/a$ &
 \multicolumn{5}{c|}{$ a^{3/2}\fhat/~Z_{stat}$}
 & $a\tilde{M}_{P}$
\\
        &           &     &                 &         &
$t_{min} = 2a $ & $t_{min} = 3a $ & $t_{min} = 4a $ & $t_{min} = 5a $
  & $t_{min} = 6a $ & \\
\hline
A1 &  0.1560 &  & 2.0 10 & 1.8
&  $0.751(35)^*$& 0.770(51)& 0.781(66)& 0.788(74)& & 0.834(28)\\
A1 &  0.1580 &  & 2.0 10 & 1.8
&  $0.751(33)^*$ & 0.771(50) & 0.785(63) & 0.791(69) & & 0.833(27)\\
\hline
A2 &  0.1560 &  & 2.0 10 & 2.6
&  $0.775(31)^*$ & 0.798(46) & 0.810(60) & 0.817(65) & & 0.833(27)\\
A2 &  0.1580 &  & 2.0 10 & 2.6
&  $0.757(31)^*$ & 0.774(47) & 0.784(61) & 0.789(65) & & 0.811(29)\\
\hline
A3 & 0.1560 & & 2.0   10 & 2
& * & 0.840(31)$^*$ & 0.919(46)  & 0.882(43) & 0.900(46) & 0.855(17) \\
A3 & 0.1620 & & 2.0   10 & 2
& * & 0.769(33)$^*$ & 0.785(41) & 0.798(43) & 0.811(41)  & 0.785(21) \\
A3 & 0.1635 & & 2.0   10 & 2
& * & 0.745(31)$^*$ & 0.756(39) & 0.766(39) & 0.780(34) & 0.768(20) \\
\hline
A3a &  0.1560 & 0.1866 &&
&  0.874(16) & 0.884(26) & 0.879(29) & 0.884(59) & & 0.837(5)\\
A3a & 0.1600 & 0.1866 &&
&  0.840(18) & 0.836(26) & 0.822(29) & 0.829(64) & & 0.792(5)\\
A3a & 0.1620 & 0.1866 &&
&  0.819(25) & 0.808(35) & 0.794(44) & 0.803(78) & & 0.772(6)\\
A3a & 0.1635 & 0.1866 &&
&  0.805(33) & 0.790(43) & 0.772(56) & 0.780(95) & & 0.758(8)\\
\hline
A4 &  0.1560 &  & 2.0 10 & 3.6
&  $0.782(22)^*$ & $0.809(35)^*$ & 0.829(43) & 0.841(50) &
0.835(54)&0.831(21)\\
A4 &  0.1580 &  & 2.0 10 & 3.6
&  $0.765(25)^*$ & 0.788(36) & 0.804(44) & 0.813(51) &  0.804(55)& 0.810(23)\\
\hline
A5 & 0.1560 & & 2.0 10 &  &
$0.777(19)^*$ & $0.809(33)^*$ & 0.854(42) & 0.834(46) & 0.840(49) & 0.841(18)\\
A5 & 0.1620 & & 2.0 10 & &
$0.740(18)^*$ & 0.768(31) & 0.781(40) & 0.790(40) & 0.793(37) & 0.792(18) \\
A5 & 0.1635 & & 2.0 10 & &
$0.719(22)^*$ & 0.760(44) & 0.745(47) & 0.748(50) & 0.745(47) & 0.779(20) \\
\hline
B1 & 0.1557 & 0.1856& & 1.9
& * & 0.635(60) & 0.618(69) & 0.590(81) & & 0.764(13)\\
B1 & 0.1574 & 0.1856& & 1.9
& * & 0.618(63) & 0.596(71) & 0.571(81) & & 0.749(15) \\
B1 & 0.1587 & 0.1856& & 1.9
& * & 0.609(72) & 0.581(81) & 0.558(98) & & 0.746(18)   \\
\hline
C1 & 0.1525 & & 4.0 50 & 3.8
&  0.313(19) & 0.322(27) & 0.326(33) & 0.325(37) &  & 0.646(28)\\
\hline
C2 & 0.1500 & & 4.0 25 & 3.2
& $0.333(16)^*$ & $0.349(22)^*$ & 0.368(31) & 0.376(50) &  & 0.704(9)\\
C2 & 0.1525 & & 4.0 25 & 3.2
& 0.351(18) & 0.356(24) & 0.364(37) & &  & 0.669(10)\\
\hline
C3 & 0.1525 &  & 4.0 50 & 4.0
& $0.345(10)^*$ & $0.365(18)^*$ & 0.379(24)  & 0.388(31) & 0.397(43)
& 0.690(18) \\
C3 & 0.1540 &  & 4.0 50 & 4.0
& $0.330(10)^*$ & $0.347(16)^*$ & 0.360(29) & 0.369(26) & 0.375(37)
& 0.665(18)\\
C3 & 0.1558 &  &4.0 50 & 4.0
& $0.316(15)^*$ & $0.334(24)^*$  & 0.346(30) & 0.354(37) & 0.365(47)
& 0.652(16)\\
\hline
C3a & 0.1525 & 0.1854 && 4.0
& * & $0.451(06)^*$ & $0.415(14)^*$ & 0.394(19) & 0.390(25)& 0.687(12) \\
C3a & 0.1540 & 0.1854 && 4.0
& * & $0.431(06)^*$ & $0.395(14)^*$  & 0.371(20) & 0.369(26) & 0.666(13)\\
C3a & 0.1550 & 0.1854 && 4.0
& * & $0.416(08)^*$ & $0.381(14)^*$ & 0.359(20) & 0.359(27) & 0.653(14)\\
\hline
C4 & 0.1525 & & 4.0   50$^\dagger$ & &
* & $0.353(14)^*$ & $0.361(16)^*$ & 0.368(18) & 0.371(19) & 0.672(14) \\
\hline
D1 & 0.1492 &  &4. 100 & 5.0
& $0.190(11)^*$ & $0.195(14)$ & 0.201(16) & 0.206(20) & 0.204(21)
& 0.541(31)\\
\hline
D2 & 0.1492 &  &4. 100 & 5.5
& $0.203(09)^*$ & $0.211(11)^*$ & 0.214(13) & 0.213(13) &
& 0.598(33)\\
D2  & 0.1506 &  & 4. 100  & 5.5
& $0.190(09)^*$ & $0.196(11)^*$ & 0.198(11) & 0.195(11) &
& 0.575(31)\\
D2  & 0.1514 &  & 4. 100  & 5.5
& $0.184(10)^*$ & $0.189(11)^*$ & 0.188(11) & 0.185(11) &
& 0.565(31)\\
\hline
D2a & 0.1492 &  &4. 100 & 5.6
& $0.188(15)^*$ & $0.189(16)^*$ & 0.190(16) & 0.190(18) & 0.190(19)
& 0.565(8)\\
\hline
\end{tabular}
}
\end{center}
\pagestyle{plain}
\newpage
\item[\bf Table 4:]
 The mass splittings $\Delta_{S}$ and $\Delta_{\Lambda}$ and the scale of
string
 breaking  $R_b$ in lattice units are given. An asterisk indicates that no
 clear plateau
 could be identified. The second column gives the parameters for the gaussian
 trial wave functions used. For all lattices smearing was applied either to
 the heavy or light quark except for lattice C4 where both quarks were smeared.
 A dagger $({\dagger})$ denotes numbers obtained not by fitting the ratio $R_S$
but
 separately the scalar and pseudoscalar correlators.

\begin{center}
\begin{tabular}{|c|c|c|c|c|c|}
\hline
latt.& $\alpha$ ~~$n$  & $\kappa$& $a\Delta_{S}$& $a\Delta_{\Lambda}$&
$R_b/a$\\
\hline
    & 2. 10 & 0.1560 &  0.431(20) & 0.57(8)*  & 7.9(3)\\
A3  & 2. 10 & 0.1620 &  0.482(92)* & 0.46(3)* & 6.9(3)\\
    & 2. 10 & 0.1635 &  0.500(130)* & 0.43(6)* & 6.7(3) \\
\hline
    & 2. 10 & 0.1560  & 0.413(21) & 0.35(9)* & 7.7(3)\\
A5  & 2. 10 & 0.1620  & 0.435(32) & *       &7.0(3) \\
    & 2. 10 & 0.1635  & 0.449(34) & *       &6.8(3) \\
\hline
    &       &  0.1525 & 0.210(12)& *& 15.3(9)\\
 C3 & 4. 50 &  0.1540 & 0.196(21) &* &14.4(9) \\
    &       &  0.1558 & 0.136(86) &* &  13.9(8)\\
\hline
C4 & 4. 50 & 0.1525 & 0.210(18) & 0.326(8) & 14.6(8)\\
\hline
D1 & 4. 100 & 0.1492 & 0.174(11)* & *  &21(3)\\
\hline
  & 4. 100 & 0.1492 & $0.149(57) ^{\dagger}$&  * &26(3)\\
D2  & 4. 100 & 0.1506 & $0.159(53)^{\dagger}$ & *&24(3)\\
  & 4. 100 & 0.1514 & $0.172(51)^{\dagger}$ & *&23(3)\\
\hline
D2a & 4. 100 & 0.1492 & 0.142(18) & 0.216(14)  &23(1)\\
\hline
\end{tabular}
\end{center}
\newpage
\item[\bf Table 5:]
Values for the constant piece of the quenched $Q\bar{Q}$ potential
(eq.~\ref{pot})
and the String tension. \\
\begin{center}
\begin{tabular}{|c|c|c|c|}
\hline
 $\beta$ & $a~V_0$ & $a^2~\sigma$ & Reference  \\
\hline
5.70 & 0.631(3) & 0.168(1) & \cite{MTC} \\
5.74 & 0.636(5) & 0.141(2) & interpolated \\
5.80 & 0.643(3) & 0.109(1) & \cite{MTC} \\
5.90 & 0.647(3) & 0.073(2) & \cite{MTC} \\
6.00 & 0.630(4) & 0.050(1) & \cite{BS} \\
6.20 & 0.614(1) & 0.0271(2)& \cite{BS} \\
6.26 & 0.607(2) & 0.0229(3)& interpolated \\
6.40 & 0.589(1) & 0.0154(1)& \cite{BS} \\
\hline
\end{tabular}
\end{center}

\item[\bf Table 6:]
The $\kappa$-values $\kappa_c$ and $\kappa_{s}$ corresponding
to zero and strange quark mass are given for each $\beta$. In the
two last columns we list the values of $M_{\rho}$ in lattice units at
 $\kappa_c$
and $\kappa_s$. The values of $\kappa_s$ are obtained using $M_{\rho}$
as a reference scale.\\
\begin{center}
\begin{tabular}{|c|c|c|c|c|}
\hline
 $\beta$ & $\kappa_c$ & $\kappa_{s}$ & $aM_{\rho}$& $aM_{\rho_s}$
  \\
\hline
5.74 & 0.16631(19) & 0.16194(15) & 0.534(11)& 0.660(8)\\
6.0 & 0.15716(14) &0.15493(12)  &0.341(15) & 0.427(10)\\
6.26 &0.15234(12) &0.15139(9)  &0.208(18) & 0.263(14)\\
\hline
\end{tabular}
\end{center}

\item[\bf Table 7:]
We give the bare decay constant $\fhat~/~Z_{stat}$
and pseudoscalar mass in lattice units extrapolated at $\kappa_c$
and at $\kappa_s$. The latter was fixed by using both
$M_{\rho}$ and the string tension (in
square brackets) as a reference scale. The subscript $u$ and $s$ denote
 quantities
evaluated at the chiral limit and at the strange quark mass respectively.\\
\begin{center}
\begin{tabular}{|c|c|c|c|c|}
\hline
 $\beta$ & $a^{3/2}\fhat _u~/~Z^{stat}$ &$a\tilde{M}_{P_u}$
& $a^{3/2}\fhat _s~/~Z^{stat}$ &$a\tilde{M}_{P_s}$ \\
\hline
5.74 & 0.768(28) & 0.730(8) & 0.814(21)[0.843(17)] &0.774(5)[0.804(5)]\\
6.0 & 0.327(14) & 0.627(8) & 0.356(13)[0.369(13)]  & 0.655(11)[0.667(10)]\\
6.26 & 0.177(11) & 0.550(30) & 0.189(12)[0.198(12)] &0.565(31)[0.577(31)] \\
\hline
\end{tabular}
\end{center}
\newpage
\item[\bf Table 8:]
The mass splittings $\Delta_{S}$ and $\Delta_{\Lambda}$ and the scale
of string breaking $R_b$ in lattice units are given
at $\kappa_c$
and at $\kappa_s$. The latter was fixed by using $M_{\rho}$ and
the string tension (in
square brackets) as a reference scale. The notation is the same as in
table~7.\\
\begin{center}
{\hspace*{-2.35 cm}
\begin{tabular}{|c|c|c|c|c|c|c|}
\hline
 $\beta$ & $a\Delta_{S_u}$& $a\Delta_{\Lambda_u}$& $R_{b_u}/a$
& $a\Delta_{S_s}$& $a\Delta_{\Lambda_s}$& $R_{b_s}/a$\\
\hline
5.74 & 0.516(29) & 0.387(58) & 6.15(19)&0.477(19)[0.450(16)]&
0.462(39)[0.512(25)]& 6.75(18)[7.16(19)]\\
6.0 & 0.211(20) & & 12.86(73)& 0.217(15)[0.223(13)]& &14.95(73)[14.39(67)]  \\
6.26 & 0.179(49) & &22(3) & 0.169(51)[0.160(52)]& & 23(3)[24(3)] \\
\hline
\end{tabular}
}
\end{center}
\normalsize

\end{enumerate}

\newpage
\noindent
{\bf Figure Captions \hfill}

\begin{enumerate}

\item[\bf Figure 1:]
Local masses as defined in eq.~(\ref{mass}) for the pseudoscalar meson in the
static approximation on an $18^3 \times 48$ lattice at $\beta=6.26$
and $\kappa=0.1492$ are plotted versus the time separation, $t$, in lattice
units. The
diamonds denote the mass from the ``local-local" correlation function.
The data denoted by $G1, G2$ and $G3$ show the local mass for smeared-smeared
correlators where the trial wave functions are combinations of a local and
a gaussian function. The rest are gaussian trial wave functions, the smallest
 being
the one with parameters $n=70, \alpha=4$ and the largest with
$n=160, \alpha=5$.\\

\item[\bf Figure 2:]
Local masses from the ``local-smeared" correlator for the pseudoscalar meson
on an $18^3 \times 48$
lattice at $\beta=6.26$ and $\kappa=0.1492$.
(i) (circles) denotes results with smearing applied to the static sink;
(ii) (diamonds)  presents results from a linear combination of a correlator
with smearing
applied to the static source and a correlator with
smearing as in (i) . In both cases smearing was done using a
 gaussian
wave function with $\alpha=4, n=100$.

\item[\bf Figure 3:]
Local masses from the ``local-smeared" correlator for the pseudoscalar meson
on an $18^3 \times 48$
lattice at $\beta=6.26$ and $\kappa=0.1492$,
with smearing using six different wave functions
listed  in table~2. The notation for the wave functions is the same as fig.~2.
The error band arising from fitting the ``smeared-smeared"
correlator for the best trial wave function is shown by the two dashed lines.

\item[\bf Figure 4:]
 Local masses from   ``local-smeared" correlators for the pseudoscalar meson
for lattices A3 and A5 at $\beta=5.74$, C3 and C4 at $\beta=6.0$ and D2a
at $\beta=6.26$ (see table~1) with the error band
arising from fitting the corresponding ``smeared-smeared" correlator. The
light quark mass is fixed so that  $M_P(l,l)/\sqrt{\sigma} \sim 4$.

\item[\bf Figure 5:]
 Local masses from  the ratio, $R_S$, of scalar to pseudoscalar
``smeared-smeared" correlator for lattices A3, A5, C3, C4 and D2a for a light
quark mass  as in fig.~4.
The time axis is in units of the string tension.

\item[\bf Figure 6:]
 Local masses from   the ratio, $R_{\Lambda}$, of scalar to pseudoscalar
``smeared-smeared" correlator for lattices A3, C3 and D2a for a light
quark mass the same as in fig.~4.
Both the time axis and the mass splitting are in units of the string tension.

\item[\bf Figure 7:]
To determine
$\kappa_c$ and  $a_{M_{\rho}}^{-1}$ at $\beta=5.74$, 6.0 and 6.26,
linear fits to $M_P^2(l,l')$ and $M_{V}(l,l')$
are shown, where
 $l,l'$ stand for fully propagating  quarks of mass $\leq 2m_{strange}$.

\item[\bf Figure 8:]
The $\rho$ mass at the chiral limit
is shown as a function of the lattice spacing both expressed in units
of the string tension. The extrapolated value at $a=0$ is obtained from
the linear fit (dashed line) to the four data points.

\item[\bf Figure 9:]
The ratio of the pseudoscalar decay constant to     the vector mass, $f_P(l,l')
/M_V(l,l')$, is plotted  as a function of $M_P^2(l,l')/M_V^2(l,l')$,
where the meaning of $l,l'$ is the same as in fig.~7. The stars denote
the experimental value for the pion and kaon.

\item[\bf Figure 10:]
$\frac{1}{Z_{stat}}\fhat \sigma^{-3/4}$ is shown vs
$L\sqrt{\sigma}$. The circles denote the results at $\beta=6.26$
the squares at $\beta=6.0$ and the diamonds at $\beta=5.74$. The light
quark mass was fixed at the same value as that in fig~4. The dashed line
is the result of fitting all data to the form $C_0-C
\exp(-1.5 L\sqrt{\sigma})$ with $C_0=3.64$ and $C=4.27$.

\item[\bf Figure 11:]
Continuum extrapolation of  $F=\frac{1}{Z_{stat}}\fhat $
at light quark mass
equal to the strange quark mass ($F_s$) and at the chiral limit ($F_u$).
The values at $a=0$ emerge  from the linear fits (dashed lines) to the data
points at
$\beta=5.74,6.0$ and 6.26.
In (a) and (b) both $F$ and $a$ are given in units of the string tension and
in (c) and (d) in units of the $\rho$ mass.

\item[\bf Figure 12:]
Continuum extrapolation of the scalar-pseudoscalar mass splitting at twice
the strange quark mass, $\Delta_{2s}$.  The values at $a=0$ result
from the linear fits (dashed lines) to the data at $\beta=5.74,6.0$
and 6.26.  In (a) we use units set by the string tension and in (b) by
the $\rho$ mass.

\item[\bf Figure 13:]
The string breaking distance $R_b \sqrt{\sigma}$ is shown vs
$a\sqrt{\sigma}$ for three light quark masses: in (a) at twice the strange
quark mass, in (b) at the strange quark mass and in (c) at the chiral limit.
Linear fits to the data at $\beta=5.74,6.0$ and 6.26 are shown by the
dashed line together with
the extrapolated values at $a=0$.

\end{enumerate}

\begin{thebibliography}{99}
\bibitem{QCDPAX}{ T. Yoshie et al.,  in {\it ``Lattice 91"}, Nucl. Phys. B
(Proc. Suppl.) 26 (1992) 281; we would like to thank Y. Iwasaki for
 communicating raw data from this computation;
this data confirms the nonlinearities that we observed.}
\end{thebibliography}
\end{document}